\documentclass[aps,prl,floatfix,twocolumn]{revtex4-1}
\usepackage{amsmath,amssymb}
\usepackage{graphicx}
\usepackage{epsfig}
\usepackage[usenames]{color}

\begin{document}

\title{Photoinduced electric currents in Bose--Einstein condensates}

\author{V.~M.~Kovalev and I.~G.~Savenko}
\affiliation{A.V.~Rzhanov Institute of Semiconductor Physics, Siberian Branch of Russian Academy of Sciences, Novosibirsk 630090, Russia\\
Center for Theoretical Physics of Complex Systems, Institute for Basic Science (IBS), Daejeon 34126, Korea}


\date{\today}

\begin{abstract}
We calculate a light-induced electric current which can occur from a Bose--Einstein condensate under the action of an external electromagnetic field with the frequency exceeding the ionization potential of the bosons, taking a system of indirect excitons as a testbed. We show that the ionization can be accompanied by the excitation of collective Bogoliubov modes. As a result, the current consists of two principal components: one regular, which has a counterpart in bosonic systems in the normal phase, and the other one specific for condensates since the photoabsorption is mediated by the emission of Bogoliubov quasiparticles. Surprisingly, the latter component soon becomes predominant with the increase of light frequency above the ionization potential.
\end{abstract}


\maketitle


\textit{Introduction.}
Photoinduced transport of charged and neutral particles is an active field of research in condensed matter physics. 
In particular, photogalvanic and photon drag effects have been widely studied first in three-dimensional metals, insulators and semiconductors~\cite{Costa, Loudon, Goff, Shalygin} and then in two-dimensional (2D) electron gas~\cite{Wieck, Luryi, Grinberg}, low-dimensional nanostructures~\cite{Mikheev}, and purely 2D materials such as graphene~\cite{Glazov, Entin, Karch} and transition metal dichalcogenides~\cite{Lee}.

The photon-drag effect (PDE)~\cite{Gibson, Danishevskii, Ivchenko} is associated with the transfer of photon momentum to the media excitations. From the classical physics perspective, PDE is a radiation pressure phenomenon. The research on the PDE has been carried out in electron and hole gases in nanostructures and recently it has been studied in gases of neutral particles~\cite{RefRPQ} such as indirect excitons in double quantum well (DQW) nanostructures~\cite{RefOurJETP, RefPDE}.
It has been demonstrated that the spectrum of photon-stimulated flux of excitons (or other bosons) can reveal a resonant~\cite{RefOurJETP} or steplike~\cite{ RefPDE} behavior as a function of the external electromagnetic (EM) field frequency.

However, there remains the problem of experimental verification of these effects since measurement of neutral particle currents (fluxes) is nontrivial. Indeed, from the experimental point of view it would be more convenient to measure the actual electric currents. 
In this work we study electric currents emerging in hybrid particle condensates exposed to an external EM field. As a lab system we consider indirect excitons in the condensed state~\cite{Butov, Fogler, Fedichkin}, however, our theory might also be applicable to exciton-polariton and atomic condensates. 

In order to reveal the existence of a condensate in an exciton-polariton system, one usually checks if there is a transition to a macroscopically coherent state which is benchmarked by a steep decrease of the emission linewidth with the simultaneous increase of the emission intensity~\cite{RefKasprzak}. Another indirect technique is the application of an external magnetic field which leads to the modulation of the emission properties of the system, thus manifesting the hybridization of the excitons and photons to form compound particles~\cite{OurNature}. In this manuscript, we are suggesting an alternative method which can be utilized to examine the existence of a condensate.

If the frequency of an external EM field exceeds the ionization potential, the photogenerated electrons and holes appear in the system and they participate in the electric currents in the direction of the momentum of light. We show that a nonzero electric current can occur only at finite value of the photon momentum tracing its origin to the PDE phenomenon. In addition to the intuitive trivial contribution, the total current also has a specific second part which is nonzero only if the system is in the condensate state since the corresponding process is mediated by the emission of Bogoliubov quasiparticles (bogolons).

\begin{figure*}[!t]
	\includegraphics[width=0.7\linewidth]{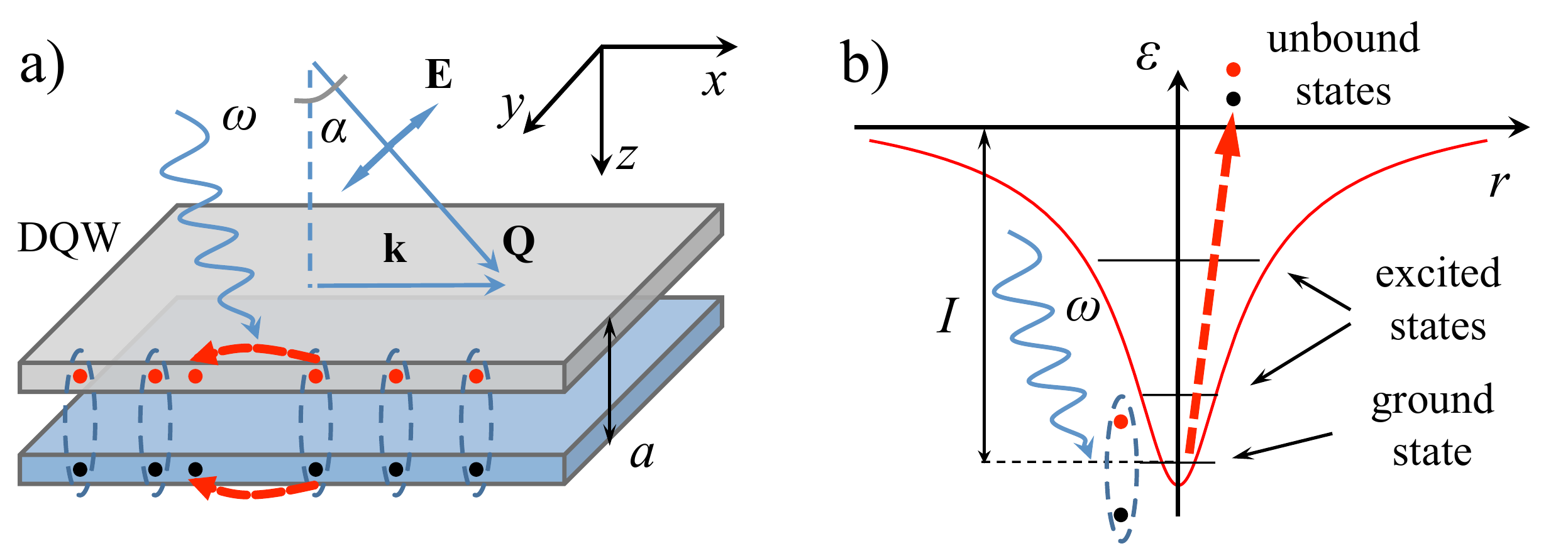}
	\caption{System schematic. 
	(a) Double quantum well (DQW) structure under an external EM field of light $\mathbf{E}$. 
	(b) Internal energy spectrum of an individual exciton. 
	Red dashed arrows show the photoionization of excitons at certain frequencies, $I$ is the ionization potential.}
	\label{Fig1}
\end{figure*}
%
%
%


\textit{Theory.} Let us consider a DWQ structure containing a dipolar exction gas (see Fig.~\ref{Fig1}). At low enough temperatures, the condensate of excitons appears in the system at the lowest state of internal exciton motion and with zero center-of-mass momentum. We illuminate the system by a monochromatic EM field with the wave vector $\textbf{Q}$ under the angle of incidence $\alpha$ with respect to the normal to the plane of the DQW. If the EM frequency exceeds the ionization potential of an individual exciton, an unbound electron-hole pair is generated as it is shown in Fig.~\ref{Fig1}(a,b). These carriers of charge have a nonzero momentum, taking part in current of photogenerated electrons and holes which reads~\cite{Entin}
\begin{equation}\label{1}
\textbf{j}=e\int (\tau_h\textbf{v}_h-\tau_e\textbf{v}_e)W(\textbf{p}_e,\textbf{p}_h)\frac{d\textbf{p}_ed\textbf{p}_h}{(2\pi)^4},
\end{equation}
where $e>0$, $\tau_{e,h}$, $\textbf{v}_{e,h}$ and $\mathbf{p}_{e,h}$ are momentum relaxation times, velocities and momenta of the electron and hole, respectively, $W(\textbf{p}_e,\textbf{p}_h)$ is a rate of electron-hole pair generation due to the exciton ionization, and we assume parabolic dispersions of the electron and hole in Eq.~(\ref{1}). It should be noted, that in general case, there exist the reverse process of electron-hole binding and interband recombination events. We consider these processes slow and thus disregard them.

Let us introduce the relative $\textbf{r}=\textbf{r}_e-\textbf{r}_h$ and center-of-mass $\textbf{R}=(m_e\textbf{r}_e+m_h\textbf{r}_h)/M$ coordinates, and the corresponding momentum operators $\textbf{p}=-\textrm{i}\partial_\textbf{R}$ and $\textbf{q}=-\textrm{i}\partial_\textbf{r}$ (thus $\textbf{p}_e=m_e\textbf{v}_e=m_e\textbf{p}/M-\textbf{q}$ and $\textbf{p}_h=m_h\textbf{v}_h=m_h\textbf{p}/M+\textbf{q}$), where $M=m_e+m_h$ is full exciton mass.
Then the current can be expressed via new variables:
\begin{equation}\label{Eq2}
\textbf{j}=e\int \left[\frac{\tau_h-\tau_e}{M}\textbf{p}+\left(\frac{\tau_h}{m_h}+\frac{\tau_e}{m_e}\right)\textbf{q}\right]W(\textbf{p},\textbf{q})\frac{d\textbf{p}d\textbf{q}}{(2\pi)^4}.
\end{equation}

The next task is to find $W(\textbf{p},\textbf{q})$ which appears in Eq.~\eqref{Eq2}. A single exciton can be described by the Hamiltonian
\begin{gather}\label{Eq3}
\hat{H}=\frac{\hat{\textbf{p}}^2}{2M}+\frac{\hat{\textbf{q}}^2}{2M}+\hat{u}(\textbf{r}),
\end{gather}
where $\hat{u}(\textbf{r})$ is the Coulomb interaction between the electron and hole. This Hamiltonian describes both the bound state of the electron and hole (exciton) and the photogenerated electron-hole pair belonging to continuous spectrum of this Hamiltonian. For the excitons located in a DQW we use $u(\textbf{r})=-e^2/4\pi\epsilon_0\epsilon\sqrt{\textbf{r}^2+a^2}$ with $a$ being the distance between the quantum wells forming the DQW. The exciton wave function $\Phi(\textbf{R},\textbf{r})$ and the wave function of the photogenerated electron-hole pair $\Psi(\textbf{R},\textbf{r})$ are given by
\begin{gather}\label{Eq4}
\Phi(\textbf{R},\textbf{r})=e^{i\textbf{p}_{ex}\textbf{R}}\phi_\eta(\textbf{r}),\\\nonumber
\Psi(\textbf{R},\textbf{r})=e^{i\textbf{p}_{eh}\textbf{R}}\psi_\textbf{q}(\textbf{r}),
\end{gather}
where the center-of-mass momenta $\textbf{p}_{ex}$ and $\textbf{p}_{eh}$ correspond to exciton and electron-hole pair; $\phi_\eta(\textbf{r})$ is the wave function of internal exciton motion, belonging to the discrete spectrum of the Hamiltonian~\eqref{Eq3} with $\eta$ indicating the set of quantum numbers of discrete exciton levels; $\psi_\textbf{q}(\textbf{r})$ describes the relative motion of electron-hole pair interacting via Coulomb potential and having a continuous spectrum.

The exciton condensate can occur at the lowest energy of internal motion which we indicate by $\eta=0$. In what follows all the energies are counted from this energy. The exciton ionization by electromagnetic field from the bound state $\eta=0$ is given by the Hamiltonian
\begin{gather}\label{Eq5}
\hat{V}=-e\int d\textbf{R}\int d\textbf{r}\hat{\Psi}^\dag(\textbf{R},\textbf{r},t)\textbf{r}\hat{\textbf{E}}(\textbf{R},t)\hat{\Phi}(\textbf{R},\textbf{r},t)=\\\nonumber
=\sum_{\textbf{p}_{eh},\textbf{q}}{\cal M}(\textbf{q})\hat{c}^\dag_{\textbf{p}_{eh},\textbf{q}}(t)\hat{a}_\textbf{k}(t)\hat{x}_{\textbf{p}_{eh}-\textbf{k}}(t),
\end{gather}
where ${\cal M}(\textbf{q})=-eE_0(2\pi)^2\int d\textbf{r}\psi^*_\textbf{q}(\textbf{r})\textbf{re}\phi_0(\textbf{r})$ and $\textbf{e}$ is the polarization vector of the EM field; $\textbf{k}$ is in-plane component of EM wave vector with absolute value $k=Q\sin\alpha$; $\hat{c},~\hat{a},~\hat{x}$ are annihilation operators of electron-hole pair, photon and exciton, respectively~\cite{OurComment}. 
%
%
In the presence of the condensate, we have
\begin{gather}\label{Eq6}
\hat{x}_{\textbf{p}}=\sqrt{n_c}\delta(\textbf{p})+u_\textbf{p}\hat{b}_\textbf{p}+v_\textbf{p}\hat{b}^\dag_{-\textbf{p}},
\end{gather}
where $n_c$ is the exciton density in the condensate and the first term $\sqrt{n_c}\delta(\textbf{p})$ describes the condensate fraction with zero center-of-mass exciton momentum $\textbf{p}=0$; the operators $\hat{b}_\textbf{p},~\hat{b}^\dag_\textbf{p}$ describe the Bogoliubov quasiparticles (collective modes of the condensate), and $u_\textbf{p},~v_\textbf{p}$
are the Bogoliubov transformation coefficients (see e.g.~\cite{RefCapture}).  

\begin{figure*}[!t]
	\includegraphics[width=0.8\linewidth]{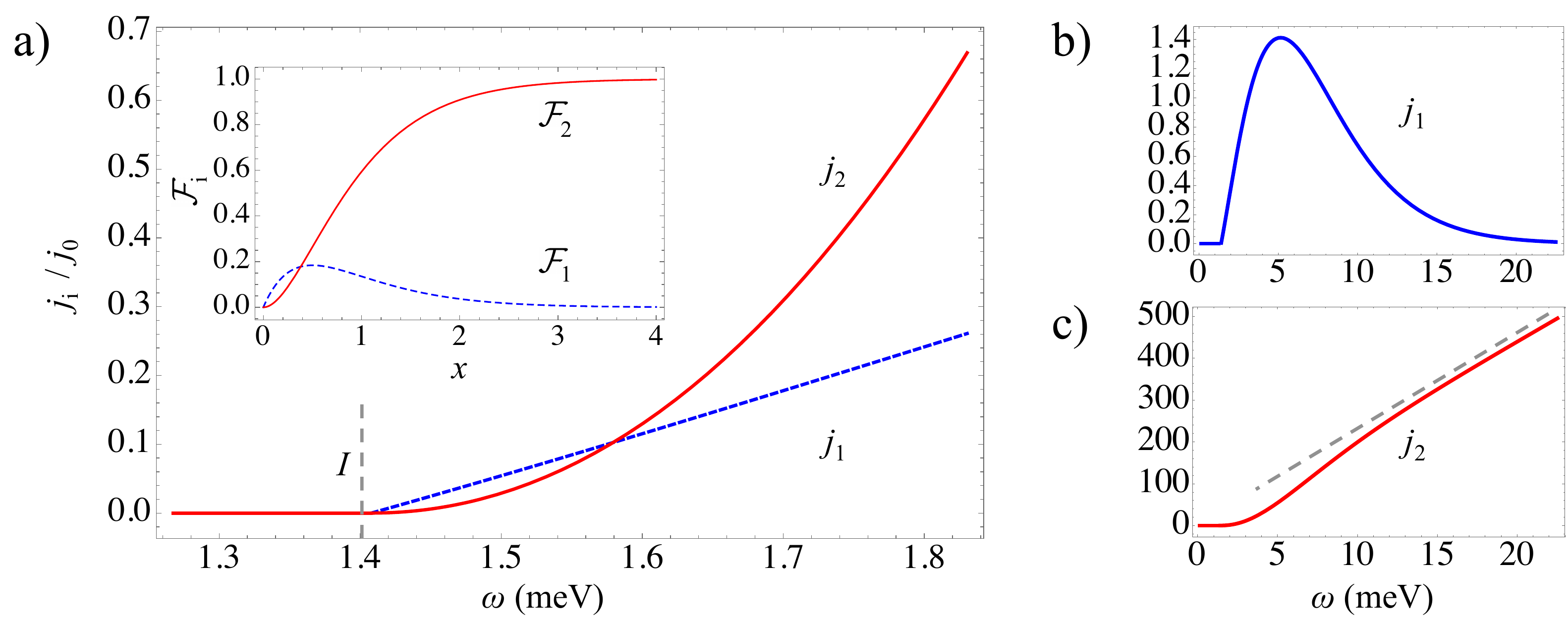}
	\caption{Components of the current in the system (in arbitrary units) at the frequencies in the vicinity of the ionization potential (a) and at higher frequencies (b,c). The component $j_1$ has an asymmetric bell-shape, saturating at high frequencies and $j_2$ has a linear asymptotic.
	Inset shows the special functions ${\cal F}_1,~{\cal F}_2$ (see text for details).}
	\label{Fig2}
\end{figure*}

Substituting Eq.~(\ref{Eq6}) into Eq.~(\ref{Eq5}) and assuming that at zero temperature the occupation numbers of Bogoliubov modes are zero, we come up with two processes of exciton ionization. The first process is given by
\begin{gather}\label{Eq8}
\hat{V}_1=\sqrt{n_c}\sum_{\textbf{p}_{eh},\textbf{q}}{\cal M}(\textbf{q})\delta(\textbf{p}_{eh}-\textbf{k})\hat{c}^\dag_{\textbf{k},\textbf{q}}\hat{a}_\textbf{k},
\end{gather}
describing creation of an electron-hole pair directly from the condensate due to the decay of the exciton. A similar process is possible in the absence of a condensate (where instead of $n_c$ in Eq.~\eqref{Eq8} there will be the difference of the occupation numbers of the ground and excited states). The corresponding probability of electron-hole pair creation can be found from the Fermi golden rule (we put $\hbar=1$ in most of places below):
\begin{eqnarray}
\label{Eq9}
W(\textbf{p}_{eh},\textbf{q})&=&2\pi n_c|\mathcal{M}(\textbf{q})
\delta(\textbf{p}_{eh}-\textbf{k})|^2\\
\nonumber
&&~~~~~~~~~\times\delta\left(\frac{\textbf{k}^2}{2M}+\frac{\textbf{q}^2}{2\mu}-\omega+I\right).
\end{eqnarray}
The second process
\begin{gather}\label{Eq10}
\hat{V}_2=\sum_{\textbf{p}_{eh},\textbf{q}}{\cal M}(\textbf{q})v_{\textbf{p}_{eh}-\textbf{k}}
\hat{c}^\dag_{\textbf{p}_{eh},\textbf{q}}(t)\hat{a}_\textbf{k}(t)\hat{b}^\dag_{-\textbf{p}_{eh}+\textbf{k}}(t)
\end{gather}
is unique for the systems containing a condensate. Here the creation of an electron-hole pair due to the exciton decay is accompanied by the emission of a bogolon with the dispersion law $\varepsilon_\textbf{p}=sp\sqrt{1+(p\xi)^2}$, where $s=\sqrt{gn_c/M}$ is a velocity of the bogolon, $g=e^2a/(\epsilon_0\epsilon)$~\cite{OurFano} is the exction-exciton interaction constant in the condensate, and $\xi=\hbar/(2Ms)$ is a healing length. 
The corresponding probability of electron-hole pair creation reads
\begin{eqnarray}
\label{Eq11}
W(\textbf{p}_{eh},\textbf{q})&=&2\pi|
\mathcal{M}(\textbf{q})|^2
|v_{\textbf{p}_{eh}-\textbf{k}}|^2\\
\nonumber
&&\times\delta\left(\frac{\textbf{p}_{eh}^2}{2M}+\frac{\textbf{q}^2}{2\mu}+\varepsilon_{-\textbf{p}_{eh}+\textbf{k}}-\omega+I\right).
\end{eqnarray}
Equations~(\ref{Eq9}) and~(\ref{Eq11}) give two contributions to the electric current defined in Eq.~\eqref{Eq2}, $\textbf{j}=\textbf{j}_1+\textbf{j}_2$. One can easily see that the second term ($\sim \textbf{q}$) under the integral in (\ref{Eq2}) gives zero  since both the probabilities~(\ref{Eq9}) and~(\ref{Eq11}) depend on the absolute value of the relative momentum $\textbf{q}$. Thus, we should only consider the first term there. 

To the lowest order with respect to the photon wave vector $\textbf{k}$, the first component of the current reads
\begin{gather}\label{12}
\textbf{j}_1=\frac{\textbf{k}en_c(\tau_h-\tau_e)\mu}{(2\pi)^2M}\theta[\omega-I]\langle|{\cal M}(\textbf{q})|^2\rangle_{q=\sqrt{2\mu(\omega-I)}},
\end{gather}
where $\theta(x)$ is the Heaviside step function and the angle averaging is defined as $\langle A\rangle=\int_0^{2\pi}A(\varphi)d\varphi/2\pi$. 
The second contribution can be found analytically in the case of linear dispersion of the Bogoliubov quasiparticles: $\varepsilon_\textbf{p}\approx sp$ and $v_p^2\approx Ms/p$, which works well in the limit $p\xi\ll 1$. After some algebra, we find:
\begin{gather}\label{13}
\textbf{j}_2=\frac{\textbf{k}e(\tau_h-\tau_e)}{(2\pi)^2}\theta[\omega-I]\int\limits_0^{\sqrt{2\mu(\omega-I)}} \langle|{\cal M}(\textbf{q})|^2\rangle
qdq.
\end{gather}

Further progress requires the knowledge of the particular form of the wave function of relative electron and hole motion both in bound (excitonic) state, $\phi_0(\textbf{r})$, and in the state of continuous spectrum, $\psi_\textbf{q}(\textbf{r})$. 
We will disregard the Coulomb interaction in the continuous part of the energy spectrum in Eq.~(\ref{Eq3}), hence $\psi_\textbf{q}(\textbf{r})$ can be approximated by the plane wave $\psi_\textbf{q}(\textbf{r})=e^{i\textbf{qr}}$, yielding
\begin{gather}\label{Eq14}
\langle|\mathcal{M}(\textbf{q})|^2\rangle=(2\pi)^4(eE_0)^2\langle|(\textbf{e}\cdot\nabla_\textbf{q})\phi_0(\textbf{q})|^2\rangle.
\end{gather}
Being the ground state eigenstate, $\phi_0(\textbf{q})$ depends on the absolute value of momentum $\textbf{q}$, $\phi_0(\textbf{q})\equiv\phi_0(q)$. Using this property, Eq.~(\ref{Eq14}) can be further expanded in the form
\begin{gather}\label{Eq15}
\langle|\mathcal{M}(\textbf{q})|^2\rangle=(2\pi)^4(eE_0)^2\left|\partial_q\phi_0(q)\right|^2\langle|(\textbf{e}\cdot \textbf{n}_\textbf{q})|^2\rangle,
\end{gather}
where $\textbf{n}_\textbf{q}=\textbf{q}/q$ is a unity vector. 
Furthermore, to find the ground state of the Hamiltonian, we expand the Coulomb interaction at small $\textbf{r}$ up to the second order terms:
\begin{gather}\label{16}
u(\textbf{r})=-\frac{e^2}{4\pi\epsilon_0\epsilon\sqrt{\textbf{r}^2+a^2}}\approx-\frac{e^2}{4\pi\epsilon_0\epsilon a}+\frac{\mu\omega_0^2r^2}{2},
\end{gather}
where $\omega_0^2=e^2/(4\pi\epsilon_0\epsilon\mu a^3)$. Within this approximation, we find
\begin{gather}\label{17}
\phi_0(\textbf{r})=\frac{1}{\ell\sqrt{2\pi}}e^{-r^2/4\ell^2}\Rightarrow\phi_0(q)=2\ell\sqrt{2\pi}e^{-q^2\ell^2},
\end{gather}
where $\ell=\sqrt{\hbar/(2\omega_0\mu)}$.
Wrapping up, we find the components of the current:
\begin{gather}\label{18}
\textbf{j}_1(\omega)=8j_0\textbf{k}\ell (\mu/M)(n_c\ell^2)\mathcal{F}_1\left(\frac{\omega-I}{\omega_0}\right),\\
\nonumber
\mathcal{F}_1(x)=x\exp(-2x)\theta(x),
\end{gather}
where 
\begin{eqnarray}
\label{Eqj0}
j_0=2(2\pi)^3e(\tau_h-\tau_e)\ell(eE_0)^2\langle|(\textbf{e}\cdot \textbf{n}_\textbf{q})|^2\rangle, 
\end{eqnarray}
%
%
%
and 
\begin{gather}\label{19}
\textbf{j}_2=j_0\textbf{k}\ell\mathcal{F}_2\left(\frac{\omega-I}{\omega_0}\right),\\
\nonumber
\mathcal{F}_2(x)=\left[1-\left(1+2x\right)\exp(-2x)\right]
\theta(x).
\end{gather}
%
%
%


\paragraph*{Results and discussion.} First of all, from Eqs.~\eqref{18} and~\eqref{19} we note that the total current is directed along the photon wave vector $\textbf{k}$ and appears in the system if the EM field frequency exceeds the exciton ionization potential $I$. Further from Eq.~\eqref{Eqj0} it follows that the total current is nonzero if the momentum relaxation times $\tau_{e,h}$ are different for electrons and holes. Let us note though, even in the case of zero total current, there will be nonzero current in each of the layers of the DQW taken separately.

The spectra of components of the total current are presented in Fig.~\ref{Fig2}. To build the plots we used parameters typical for GaAs heterostructures: $\epsilon=12.5$, $\mu=0.058m_0$, $a=20$ nm. Since our theory is only applicable to a dilute exciton gas, we put $n_c\ell^2=0.1$. The ionization potential is found as $I=e^2/(4\pi\epsilon_0\epsilon a)-\hbar\omega_0$.

The factor $\langle|(\textbf{e}\cdot \textbf{n}_\textbf{q})|^2\rangle$ in Eq.~\eqref{Eqj0} is determined by the polarization of the EM field. For linear polarization when the electric field lies in the $xz$ plane, the total current is proportional to $\sin\alpha\cos^2\alpha$. In the case of a circular polarization of the EM field, the current is proportional to $\sin\alpha(\cos^2\alpha+1)$.

From Eq.~(\ref{18}) and Eq.~(\ref{19}) and Fig.~\ref{Fig2} we see that at small frequencies $0<\omega-I\ll\omega_0$, the first term of the current $\textbf{j}_1$, associated with the direct ionization of condensed excitons, gives the main contribution. On the contrary, if $\omega-I\gg\omega_0$, the second term $\textbf{j}_2$ describing the exciton ionization accompanied by the bogolon-emitted processes exceeds $\textbf{j}_1$. Indeed, if $\omega-I\gg\omega_0$, the function ${\cal F}_1$ in Eq.~(\ref{18}) is damped exponentially, whereas in Eq.~(\ref{19}) $\mathcal{F}_2(\omega)\rightarrow 1$  (see Inset in Fig.~\ref{Fig2}). Therefore by measuring the spectrum of the electric current in the system, one can judge the existence of the condensate.



\paragraph{Conclusions.} We have studied photoinduced electric current which can occur in a system containing a Bose-Einstein condensate of complex particles exposed to an external electromagnetic field with the frequency exceeding the ionization potential. We considered indirect excitons as an example. We shown that there are two principal processes which occur in the system, one of which is accompanied by the excitation of collective Bogoliubov modes in the condensate and becomes predominant with the increase of frequency above the ionization potential. As a result, the electric current has a non-monotonous behavior allowing us to monitor the formation of a condensate.


\paragraph*{ Acknowledgments.} We acknowledge the stimulating discussion with M.~Entin. This research has been supported by the Russian Science Foundation (Project No.~17-12-01039) and the Institute for Basic Science in Korea (Project No.~IBS-R024-D1).



\begin{thebibliography}{50}
\bibitem{OurComment} It should be mentioned, that ${\cal M}(\mathbf{q})$ has a dimensionality energy$\cdot$distance$^{-1}$ due to the normalization used in Eq.~\eqref{Eq4}.
\end{thebibliography}
\end{document}